\documentclass[a4paper,twoside]{article}
\usepackage[applemac]{inputenc}
\usepackage{a4wide,amsmath,amssymb,amsthm,eucal}
\title{Einstein-Dirac theory on gauge-natural bundles}
\author{Paolo Matteucci\thanks{Faculty of Mathematical Studies, University of Southampton, 
Highfield, Southampton SO17 1BJ, England (UK). E-mail address: p.matteucci@maths.soton.ac.uk.}}
\date{14$^{\text{th}}$ September 2003}
\catcode`\£=\active
\newcommand{£}{\pounds\ida}
\makeatletter
\let\@polishl\l
\let\@polishL\L
\renewcommand{\l}{\ifmmode\lambda\else\@polishl\fi}
\renewcommand{\L}{\ifmmode\mathcal{L}\else\@polishL\fi}
\renewenvironment{proof}[1][\proofname]{\par
  \pushQED{\qed}%
  \normalfont \topsep6\p@\@plus6\p@\relax
  \trivlist
  \item[\hskip\labelsep
        \itshape
    #1\@addpunct{\textup{.}}]\ignorespaces
}{%
  \popQED\endtrivlist\@endpefalse
}
\renewenvironment{thebibliography}[1]
     {\section*{\refname
        \@mkboth{\MakeUppercase\refname}{\MakeUppercase\refname}}\small%
      \list{\@biblabel{\@arabic\c@enumiv}}%
           {\settowidth\labelwidth{\@biblabel{#1}}%
            \leftmargin\labelwidth
            \advance\leftmargin\labelsep
            \@openbib@code
            \usecounter{enumiv}%
            \let\p@enumiv\@empty
            \renewcommand\theenumiv{\@arabic\c@enumiv}}%
      \sloppy\clubpenalty4000\widowpenalty4000%
      \sfcode`\.\@m}
     {\def\@noitemerr
       {\@latex@warning{Empty `thebibliography' environment}}%
      \endlist}
\newcommand{\leftsideset}[2]{%
  \@mathmeasure\z@\displaystyle{#2}%
  \global\setbox\@ne\vbox to\ht\z@{}\dp\@ne\dp\z@
  \setbox\tw@\box\@ne
  \@mathmeasure4\displaystyle{\copy\tw@#1}%
  \@mathmeasure6\displaystyle{#2}%
  \dimen@-\wd6 \advance\dimen@\wd4 \advance\dimen@\wd\z@
  \hbox to\dimen@{}\mathord{\kern-\dimen@\box4\box6}%
}
\def\label@in@display{%
    \ifx\df@label\@empty\else
        \relax
    \fi
    \gdef\df@label
}
\makeatother

\DeclareMathOperator{\diag}{diag}

\def\inn{\mathbin{\hbox to 6pt{%
                  \vrule height0.4pt width5pt depth0pt
                  \kern-.4pt
                  \vrule height6pt width0.4pt depth0pt\hss}}}

\newcommand{\thorn}{\hbox{\textsf{\setbox0=\hbox{l}\copy0\kern-\wd0 p}}}
\newcommand{\bs}{\boldsymbol}
\newcommand{\A}{{\textstyle\bigwedge}}
\newcommand{\Ad}{\mathrm{Ad}}
\newcommand{\ag}{\mathfrak{a}}
\renewcommand{\ast}{{\displaystyle*}}
\newcommand{\Aut}{\mathrm{Aut}}

\newcommand{\bg}{\mathfrak{b}}

\newcommand{\C}{\mathbb{C}}

\newcommand{\cd}{\nabla\ida}
\newcommand{\cf}{\protect\emph{cf.{}}}
\newcommand{\Cl}{\CMcal{C}\ell}

\renewcommand{\d}{\mathrm{d}}
\newcommand{\D}{\mathcal{D}}
\newcommand{\de}{\partial}
\newcommand{\del}{\delta}
\newcommand{\dH}{\d_{\mathrm{H}}}

\def\dual(#1,#2){\langle{#1},{#2}\rangle}

\newcommand{\E}{\mathbb{E}}

\newcommand{\F}{\mathcal{F}}
\newcommand{\FF}{\mathbb{F}}
\newcommand*{\ftimes}[1]{\times_{#1}}

\newcommand{\ga}{\gamma}
\newcommand{\Ga}{\Gamma}

\newcommand{\gams}{$\gamma$~matrices}
\newcommand{\Gdash}{$G$\nobreakdash-\hspace{0pt}}

\newcommand{\Gkdash}{$G^k_m$\nobreakdash-\hspace{0pt}}
\newcommand{\gl}{\mathfrak{gl}}
\newcommand{\GL}{\mathrm{GL}}
\newcommand{\hga}{{\hat\gamma}}
\newcommand{\hrho}{{\hat\rho}}
\newcommand{\id}{\mathrm{id}}
\newcommand*{\ida}[2]{\ifx#1^{}^{#2}
                      \else\ifx#1_{}_{\!#2}
                      \else\errmessage{Sub/Superscript token missing}\fi
                      \fi}

\newcommand{\LD}{\L_{\mathrm{D}}}
\newcommand{\LH}{\L_{\mathrm{EC}}}

\newcommand{\mdash}{$m$\nobreakdash-\hspace{0pt}}

\newcommand{\onedash}{$1$\nobreakdash-\hspace{0pt}}
\newcommand{\pc}{Poincaré-Cartan}
\newcommand{\pdash}{$p$\nobreakdash-\hspace{0pt}}

\newcommand{\R}{\mathbb{R}}

\newcommand{\so}{\mathfrak{so}}
\newcommand{\SO}{\mathrm{SO}}
\newcommand{\SOdash}{$\SO(1,3)^e$\nobreakdash-\hspace{0pt}}
\newcommand{\spin}{\mathfrak{spin}}
\newcommand{\Spin}{\mathrm{Spin}}
\newcommand{\Spindash}{$\Spin(1,3)^e$\nobreakdash-\hspace{0pt}}

\newcommand{\thbE}{\bar{\E}'}
\newcommand{\thcd}{\leftsideset{^\theta}{\nabla}\ida}
\newcommand{\thE}{\E'}
\newcommand{\thG}{\leftsideset{^\theta}{\!G}\vphantom{G}}
\newcommand{\thL}{\leftsideset{^\theta}{\!\L}}
\newcommand{\thLD}{\leftsideset{^\theta}{\!\LD}}
\newcommand{\thLH}{\L_{\mathrm{EH}}}
\newcommand{\thomega}{\leftsideset{^\theta}{\!\omega}\vphantom{\omega}}
\newcommand{\thT}{\leftsideset{^\theta}{T}\vphantom{T}}
\newcommand{\thTheta}{\leftsideset{^\theta}{\!\Theta}\vphantom{\Theta}}

\newcommand{\twodash}{$2$\nobreakdash-\hspace{0pt}}
\newcommand{\U}{\mathrm{U}}
\newcommand{\Ups}{\Upsilon}
\newcommand{\VXi}{{\check\Xi}}

\newcommand{\VxiK}{\check\xi_{\mathrm{K}}{}}

\newcommand{\WGdash}{$W^{k,h}_mG$\nobreakdash-\hspace{0pt}}

\newcommand{\Z}{\mathbb{Z}}

\renewcommand{\(}{\left(}
\renewcommand{\)}{\right)}
\renewcommand{\[}{\left[}
\renewcommand{\]}{\right]}
\renewcommand{\geq}{\geqslant}
\renewcommand{\leq}{\leqslant}
\newtheorem{prop}{Proposition}[section]

\newtheorem{thm}{Theorem}[section]
\theoremstyle{definition}
\newtheorem{dfn}{Definition}[section]
\newtheorem{rem}{Remark}[section]
\numberwithin{equation}{section}

\begin{document}
\hyphenation{Frau-en-die-ner Gia-chet-ta Hil-bert rai-fear-taigh}

\maketitle
\begin{abstract}
We present a clear-cut example of the importance of the functorial approach of gauge-natural
bundles and the general theory of Lie derivatives for classical field theory, where the sole
correct geometrical formulation of Einstein (-Cartan) gravity coupled with Dirac fields gives
rise to an unexpected indeterminacy in the concept of conserved quantities.\\ \\
    \emph{Math.\ Subj.\ Class.}\ (2000): primary: 83C40;	secondary: 53C80, 83D05\\
    \emph{Keywords}: Einstein-Cartan-Dirac theory; conserved quantities; gauge-natural bundles; Lie derivative
of spinor fields
\end{abstract}

\pagestyle{myheadings}
\markboth{\itshape Paolo Matteucci}%
         {\itshape Einstein-Dirac theory on gauge-natural bundles}
\thispagestyle{plain}

\section*{Introduction}

It is commonly accepted nowadays that the appropriate mathematical arena for classical field
theory is that of fibre bundles or, more precisely, of their jet prolongations
\cite{atiyah79,trautman80,saunders89,gms97}. What is less often realized or stressed is that,
in physics, fibre bundles are always considered \emph{together with some special class of
morphisms}, i.e.\ as elements of a particular \emph{category}. The category of \emph{natural
bundles} was introduced about thirty years ago \cite{nijenhuis72,salvioli72,terng78,kms93} and
proved to be an extremely fruitful concept in differential geometry \cite{pt77,et79,km87,
eck86,luciano88}. But it was not until recently, when a suitable generalization was introduced,
that of \emph{gauge-natural bundles} \cite{eck81,kms93}, that the relevance of  this functorial
approach to physical applications began  to be clearly perceived
\cite{fffg98,gmff00,fatibene99,fafr98,gmv01,ffp01,fafr01}.

Indeed, \emph{every} classical field theory can be regarded as taking place on some jet
prolongation of some gauge-natural (vector or affine) bundle associated with some principal
bundle over a given base manifold \cite{eck81,kms93,fatibene99}.

On the other hand, it is well known that one of the most powerful tools of Lagrangian field
theory is the so-called ``Noether theorem'' \cite{noether18,kms93,gms97}. It turns
out that, when phrased in modern geometrical terms, this theorem crucially involves the concept
of a Lie derivative, and here is where the aforementioned functorial approach is not only
useful, but also \emph{intrinsically unavoidable}. On relying on the general theory of Lie
derivatives \cite{trautman72,jk82,kms93,fffg96,gm02}, it is easy to see that the concept of Lie
differentiation is, crucially, a \emph{category-dependent} one, and it makes a \emph{real} difference
in taking the Lie derivative of, say, a vector field if one regards the tangent bundle as a
purely natural bundle or, alternatively, as a more general gauge-natural bundle associated with
some suitable principal bundle (\cf~\cite{gm02}).

In this paper, we show that this functorial approach is \emph{essential} for a correct
geometrical formulation of the Einstein (-Cartan) -Dirac theory and, at the same time, yields an
unexpected \emph{indeterminacy} in the concept of conserved quantities. In the
Einstein-Cartan-Dirac case, such an indeterminacy \emph{can be regarded} as the well-known
indeterminacy which occurs in gauge theory \cite{noether18,gms97,fatibene99,bb01}, although
there are serious conceptual risks involved in dismissing this ``metric-affine'' theory of
gravitation as a standard ``gauge theory'' \cite{trautman80,gms97}. This is certainly not the
case, though, for the Einstein-Dirac theory proper, which can by no means be viewed as such. We
shall show that, in both cases, this indeterminacy actually arises from the very fact that, when
coupled with Dirac fields, Einstein's general relativity can no longer be regarded as a purely
natural theory because, in order to incorporate spinors, one \emph{must} enlarge the class of
morphisms of the theory.
 
Indeed, it is well-known that there are no representations of the group $\GL(4,\R)$ of the
automorphisms of~$\R^4$ which behave like spinors under the subgroup of Lorentz transformations.
Therefore, if one aims at considering the coupling between general relativity and fermionic
fields, one is forced  to resort to the so-called ``tetrad formalism'' (\cf, e.g.,
\cite{weinberg72}). Yet, there seems to have been a widespread misunderstanding of the full
mathematical (and physical) significance of this.
Leaving all the technicalities to the later sections, it will suffice here to
sketchily recall how the concept of a tetrad is usually introduced.

On relying on the ``principle of equivalence'', which mathematically is tantamount to the simple
statement that every manifold is locally flat, at every point~$\tilde x$ of space-time one
can erect a set of coordinates $(X^a)$ that are \emph{locally inertial} at~$\tilde x$. The
components of the metric in any general \emph{non-inertial} coordinate system are
then\footnote{Here and in the sequel both Latin and Greek indices range from~$0$ to~$3$.}
\begin{equation}
g_{\mu\nu}(x) = \theta^a\ida_\mu(x)\theta^b\ida_\nu(x)\eta_{ab},
\tag{$i$}
\label{eq:gtt}
\end{equation}
where $\lVert\eta_{ab}\rVert:=\diag(1,-1,-1,-1)$ and
\begin{equation}
\theta^a\ida_\mu(\tilde x):=\left.\frac{\de X^a(x)}{\de x^\mu}\right\rvert_{x=\tilde x},
\tag{$ii$}
\label{eq:tet-dfn}
\end{equation}
Thus, if we change our general non-inertial coordinates from $(x^\mu)$ to $(x'{}^\mu)$,
$\theta^a\ida_\mu$ will change according to the rule
\begin{equation}
\theta^a\ida_\mu \mapsto \theta'{}^a\ida_\mu = \frac{\de x^\nu}{\de x'{}^\mu}\theta^a\ida_\nu.
\tag{$iii$}
\label{eq:tet-tn}
\end{equation}
Therefore, $(\theta^a\ida_\mu)$ must be regarded as the components of \emph{four
\onedash forms}~$(\theta^a)$, \emph{not} of a single tensor field~$\theta$. This set of four \onedash forms
is what is known as a \emph{tetrad}.

At this stage, the Latin index~$a$ is just a ``label'' and, for any~$a$, $\theta^a$ is indeed a
natural object, i.e., roughly speaking, a section of a fibre bundle over the space-time
manifold~$M$ such that every coordinate change on the fibre is induced by some coordinate change
on~$M$. But the reason why a tetrad was introduced in the first place is precisely that we then
wanted to ``switch on'' that Latin index in order to incorporate spinors into our formalism.
This means that
$\theta^a\ida_\mu$ will have to additionally change according to the rule
\begin{equation}
\theta^a\ida_\mu(x) \mapsto L^a\ida_b(x)\theta^b\ida_\mu(x),
\tag{$iv$}
\label{eq:tet-tg}
\end{equation}
where~$L(x)$ is the (space-time-dependent) Lorentz transformation induced (modulo a sign) by a
given spinorial transformation~$S$ under the group epimorphism $\Lambda\colon\Spin(1,3)^e
\to\SO(1,3)^e$. 

This is precisely the point that has been too often overlooked. Unlike~\eqref{eq:tet-tn},
transformation law~\eqref{eq:tet-tg} does \emph{not} descend from definition~\eqref{eq:tet-dfn},
but is a requirement we have imposed \emph{a posteriori}. In other words, we have \emph{changed
the definition} of~$\theta^a\ida_\mu$ in such a way that now
$(\theta^a\ida_\mu)$ must be regarded as the components of a \emph{non-natural} object~$\theta$.

There is another important point that has been traditionally overlooked, which is of pre-eminent
physical significance. Recall, indeed, that spinor fields can be defined on a manifold~$M$ only
if $M$ admits a ``spin structure''. Now, the standard definition of a spin structure involves
\emph{fixing a metric} on~$M$, a framework which is certainly well-suited to a situation in which
the  gravitational field is considered \emph{unaffected} by spinors, but is otherwise unable to
describe the complete interaction and feedback between gravity and spinor fields
\cite{vdheuvel94,slawianowski96}.  To this end, the concept of a \emph{free spin structure} must
be introduced.

Ultimately, the solution to both the aforementioned problems lies in suitably defining the bundle
of which~$\theta$ is to be a section. This leads to the concept of a \emph{spin-tetrad}, which
turns out to be a gauge-natural object \cite{fffg98,gmff00,gmv01}. 

%

In the hope of making the paper understandable to both physicists and mathematicians, we have
tried to make it as self-contained as possible. Its structure is the following: in
\S\ref{sec:prelim} preliminary notions on jets, principal bundle morphisms and Clifford algebras
are recalled for the main purpose of fixing our notation. In \S\ref{sec:gnb} gauge-natural
bundles and a generalized notion of a Lie derivative are introduced. In \S\ref{sec:nt} a version
of Noether's theorem suitable for gauge-natural bundles is given. In \S\ref{sec:st} the concepts
of a spin-tetrad and a spin-connection are defined. Finally, in \S\ref{sec:ecd} we briefly
recall the Lagrangian formulation of the Einstein (-Cartan) -Dirac theory and, on applying the
theory of conserved quantities, find a general superpotential, which essentially proves the
aforementioned indeterminacy of any conserved charge associated with the gravitational field.


\section{Preliminaries and notation}\label{sec:prelim}

Throughout the paper, we shall assume that all maps are smooth, i.e.\ of class $C^\infty$, and
all manifolds are real, finite-dimensional, Hausdorff, second-countable and, hence,
paracompact. 

\subsection{Jets}\label{ssec:jets}
We assume that the reader is familiar with the standard concepts and language of differential
geometry on fibred manifolds, jet prolongation theory and calculus of variations on fibred
manifolds. Standard references are \cite{saunders89,kms93,gms97}.

Let $\pi\colon B\to M$ be a fibred manifold.
We shall denote by $\Ga(B)$ the space of all its (local) sections and set $m:=\dim M$ and
$n :=\dim B-m$. On~$B$ we shall use fibred charts $(V,x^\l,y^\ag)$, $\l=0,\dots,m-1$,
$\ag=1,\dots,n$, where~$V$ is an open subset of~$B$ projecting on the domain~$U$ of a chart
$(U,x^\l)$ of~$M$. If $\pi'\colon M'\to B'$ is another fibred manifold, by a 
\emph{fibred} (\emph{manifold}) \emph{morphism} between~$B$ and~$B'$ we shall mean a pair
$(\varphi,\Phi)$, where $\varphi\in C^{\infty}(M,M')$, $\Phi\in C^{\infty}(B,B')$ and 
$\pi'\circ\Phi=\varphi\circ\pi$. In particular, a \emph{base-preserving} (\emph{fibred})
\emph{morphism} from~$B$ to~$B'$ will be a fibred morphism between~$B$ and~$B'$ for which
$M'\equiv M$ and $\varphi\equiv\id_M$.

Recall that two curves $\gamma,\delta\colon\R\to M$ are said to have \emph{contact of order~$k$
at zero} if, for every smooth function $\varphi\colon M\to\R$, all derivatives up to
order~$k$ of the difference $\varphi\circ\gamma-\varphi\circ\delta$ vanish at~$0\in\R$. Two maps
$f,g\colon M\to N$ are then said to determine the same \emph{$k$-jet} at $x\in M$ if, for every
curve $\gamma\colon\R\to M$ with $\gamma(0)=x$, the curves $f\circ\gamma$ and $g\circ\gamma$ have
contact of order~$k$ at zero, and we shall write $j^k_x f=j^k_x g$.
In particular, let $\pi\colon B\to M$ be a fibred manifold: the set~$J^k\!B$ of all $k$-jets of
its local sections has a natural topology of a fibred manifold over~$M$, denoted by $\pi^k\colon
J^k\!B\to M$ and called the \emph{$k$-th order jet prolongation} of~$\pi\colon B\to M$. If
$\pi\colon B\to M$ is a bundle, so is $\pi^k\colon J^k\!B\to M$. Its holonomic sections are called
\emph{$k$-th order jet prolongations} of sections of $\pi\colon B\to M$ and will be denoted by
$j^k\sigma$ for any given $\sigma\in\Gamma(B)$. The adapted fibred chart on~$J^k\!B$ induced by
the chart $(V,x^\l,y^\ag)$ on~$B$ will be denoted by
$(J^kV,x^\l,y^\ag\ida_{\bs\mu})$, where $\bs\mu$ is a multi-index of length
$\lvert\bs\mu\rvert$ such that $0\leq\lvert\bs\mu\rvert\leq k$. Moreover, we shall set
$\de_\l:=\de/\de x^\l$, $\de_\ag\ida^{\bs\mu}:=\de/\de
y^\ag\ida_{\bs\mu}$ and $J^0\!B := B$, as customary.

Let $\F(B)$ denote the ring of smooth, real-valued functions over~$B$. We define an
$\F(B)$-submodule~$\Omega^p_0(B)$ of the module~$\Omega^p(B)$ of \pdash forms over~$B$ according to
the following prescription: $\omega\in\Omega^p_0(B)$ iff $\Ups\inn\omega=0$ for any
\emph{vertical vector field}~$\Ups$ on~$B$ (i.e.\ any vector field~$\Ups$ such that
$T\pi\circ\Ups=0$), `$\inn$' denoting the interior product. The elements of
$\Omega^p_0(B)$ are called \emph{horizontal \pdash forms} (on~$B$) and are in one-to-one
correspondence with the base-preserving morphisms from~$B$ to~$\A^pT^*\!M$. Furthermore, we say
that $\omega\in\Omega^p(J^k\!B)$ is a \emph{contact \pdash form} iff $(j^k\sigma)^\ast\omega=0$ for
all local section~$\sigma$ of~$B$.

Now, let $\pi^k_h\colon J^k\!B\to J^h\!B$, $k\geq h$, be the canonical projection. Recall that
for any \pdash form $\omega\in\Omega^p(J^k\!B)$, $p\leq m$, there exists a unique invariant
decomposition
\begin{equation*}
(\pi^{k+1}_k)^\ast\omega = h(\omega) + k(\omega)
\end{equation*}
into its \emph{horizontal part} $h(\omega)\in\Omega^p_0(J^{k+1}\!B)$ and \emph{contact part}
$k(\omega)\in\Omega^p(J^{k+1}\!B)$. In particular, given a function $f\in\F(J^k\!B)$, in any
fibred chart we have $h(\d f) := \d_\mu f\,\d x^\mu$, where $\d_\mu f$ denotes the \emph{formal}
or \emph{total} (\emph{coordinate}) \emph{derivative} of~$f$, defined by requiring
\begin{equation*}
    (\d_{\mu}f)\circ j^{k+1}\sigma =\de_{\mu}(f\circ j^{k}\sigma)
\end{equation*}
for all $\sigma\in\Gamma(B)$. We shall also write $\d_{\bs\mu}$, $1\leq\lvert\bs\mu\rvert\leq
k$, for $\d_{\mu_k}\dotsm\d_{\mu_1}$. Analogously, we define the \emph{horizontal
differential}~$\dH\omega$ of any form $\omega\in\Omega^{p}(J^{k}\!B)$ by requiring
\begin{equation*}
    (j^{k+1}\sigma)^{\ast}\dH\omega =(j^{k}\sigma)^{\ast}\d\omega
          \equiv\d[(j^{k}\sigma)^{\ast}\omega]
\end{equation*}
for all $\sigma\in\Gamma(B)$.

\subsection{Principal bundle morphisms}
\label{ssec:pbm}

For the reader's convenience, we recall herein some basic ideas on principal bundle
morphisms (\cf, e.g., \cite{kn63}).

Let $P(M,G)$ be a principal (fibre) bundle. A (\emph{principal}) \emph{automorphism} of~$P$
is a \Gdash equivariant diffeomorphism of~$P$ onto itself, i.e.\ a diffeomorphism
$\Phi\colon P\to P$ such that $\Phi(u\cdot a) =\Phi(u)\cdot a$  for all $u\in
P$ and $a\in G$, `$\cdot$' denoting the canonical right action of~$G$ on~$P$. We shall 
denote by $\Aut(P)$ the group of all automorphisms of~$P$.

Now, let $\Xi$ be a vector field on~$P$ generating a one-parameter group $\{\Phi_{t}\}$. Then, 
$\Xi$ is called \emph{\Gdash invariant} if $\Phi_{t}$ is an automorphism of $P$ for all
$t\in\R$.

Owing to \Gdash equivariance, each automorphism~$\Phi\in\Aut(P)$ induces a unique diffeomorphism
$\varphi \colon M\to M$ such that $\pi \circ \Phi = \varphi \circ \pi$, $\pi$ denoting the
canonical projection of~$P$ on~$M$. Then it follows immediately that every \Gdash invariant
vector field~$\Xi$ on~$P$ is \emph{projectable} over a unique vector field~$\xi$ on the base
manifold~$M$, i.e.\ $T\pi\circ\Xi=\xi\circ\pi$.

\subsection{Clifford algebra, \gams\ and spin group}\label{sec:cliff}

The \emph{Clifford algebra} $\Cl(V)$ on a (real) vector space~$V$ equipped with a scalar product
$(u,v)\mapsto g(u,v)$ is an associative algebra such that there exists a linear map~$\gamma$
from~$V$ into a subset of~$\Cl(V)$ generating~$\Cl(V)$ and satisfying the property
\begin{equation*}
\ga(u)\ga(v) +\ga(v)\ga(u) = -2g(u,v)e,
\end{equation*}
$e$ denoting the unit element of~$\Cl(V)$. The Clifford algebra on an \mdash dimensional vector
space has dimension~$2^m$. It can be realized by an algebra of linear maps of a complex vector
space of dimension~$2^{I(m/2)}$ into itself, $I(m/2)$ denoting the integral part of~$m/2$.

By \emph{\gams} we shall mean a set of~$m$ such linear maps, represented by
matrices, associated with the vectors of an orthonormal frame of~$V$.

If we denote by~$(\eta_{ab})$ the components of~$g$ in such a frame, then the
\gams, which we shall denote by~$(\ga_a)$, satisfy the fundamental relation  
\begin{equation}
    \ga_a\ga_b +\ga_b\ga_a =-2\eta_{ab},
\label{eq:cp}
\end{equation}
where the identity matrix is implied on the right-hand side. We shall also define
\begin{equation*}
    \ga_{a_1\dots a_k} := \frac1{k!}\ga_{[a_1}\dotsm\ga_{a_k]}.
\end{equation*}
In fact, it turns out that we need to consider only antisymmetrized products. This is because, on
applying~\eqref{eq:cp} iteratively, we find
\begin{gather*}
\ga_a\ga_b = \ga_{ab} -\eta_{ab},  \notag  \\
\ga_a\ga_b\ga_c = \ga_{abc} -\eta_{ab}\ga_c -\eta_{bc}\ga_a +\eta_{ca}\ga_b,
\end{gather*}
and so forth. Moreover, substituting the former into the latter relation yields
\begin{equation}
\ga_{ab}\ga_c +\ga_c\ga_{ab} = 2\ga_{abc},
\label{eq:gabc}
\end{equation}
an identity which will prove useful later on.

Henceforth, $V$ will be assumed to be 4-dimensional and $g$ will have signature $(1,3)$.
Therefore, $\lVert\eta_{ab}\rVert = \diag(1, -1, -1, -1)$. Furthermore, it can be shown that the
\gams\ satisfy the following [anti] Hermiticity properties:
\begin{equation}
\ga_a^\dag =
\begin{cases}
-\ga_a &\text{if $a=0$},\\
\ga_a  &\text{if $a=1,2,3$},
\end{cases}
\label{eq:hp}
\end{equation}
`$\dag$' denoting transposition and complex conjugation. From~\eqref{eq:cp} and~\eqref{eq:hp} it
follows immediately that
\begin{equation*}
\ga_0\ga_a\ga_0^{-1} = -\ga_a^\dag.
\end{equation*}

Finally, by the \emph{spin group} $\Spin(1,3)$ we shall mean the subgroup of $\GL(4,\C)$
consisting of those elements~$S$ such that there exists an $L\in\SO(1,3)$ satisfying
\begin{subequations}\label{eq:sg}
\begin{equation}
S\ga_a S^{-1}= L_a\ida^b\ga_b, 
\label{eq:sg1}
\end{equation}
$L\equiv\lVert L_a\ida^b\rVert$, and such that
\begin{equation}
\det(S)=1.
\label{eq:sg2}
\end{equation}
\end{subequations}
Relations~\eqref{eq:sg} define an epimorphism from $\Spin(1,3)$ onto $\SO(1,3)$. It can be shown
that $\Spin(1,3)$ [$\Spin(1,3)^e$] is the twofold covering of $\SO(1,3)$ [$\SO(1,3)^e$], the
superscript~$e$ denoting the connected component with the unit. In particular, $\Spin(1,3)^e$ is
simply connected.

\section{Gauge-natural bundles}\label{sec:gnb}

The concept of a gauge-natural bundle was originally introduced by Eck in~\cite{eck81}. 
As we have applications in mind, in this section we shall follow a constructive approach along
the lines of \cite{kms93}, notably \S15 and \S52.4. An (equivalent) axiomatic formulation can be
found again in \cite[Chapter~XII]{kms93}.

\begin{dfn}
The set
\begin{equation*}
\{\,j^k_0\alpha \mid \alpha\colon\R^{m} \to \R^{m}\text{, }
\alpha(0)=0 \text{, locally invertible}\,\}
\end{equation*}
equipped with the jet composition $j^k_0\alpha\circ j^k_0\alpha':=j^k_0(\alpha\circ\alpha')$ is a
Lie group called the \emph{$k$-th differential group} and denoted by~$G^k_m$.
\end{dfn}

\noindent
For $k=1$ we have, of course, the identification $G^1_m\cong\GL(m,\R)$.

\begin{dfn}
Let $M$ be an $m$-dimensional manifold. The principal bundle over~$M$ with group $G^k_m$
is called the \emph{$k$-th order frame bundle} over~$M$ and will be denoted by $L^k\!M$.
\end{dfn}

\noindent
For $k=1$ we have, of course, the identification $L^1\!M\cong LM$, where $LM$ is the usual
(principal) \emph{bundle of linear frames} over~$M$ (\cf, e.g., \cite{kn63}).

\begin{dfn}
Let $G$ be a Lie group.
Then, by the \emph{space of $(m,h)$-velocities} of~$G$ we shall mean the set
\begin{equation*}
T^h_mG:=\{\,j^h_0a\mid a\colon\R^{m} \to G\,\}.
\end{equation*}
\end{dfn}

\noindent
Thus, $T^h_mG$ denotes the set of $h$-jets with ``source'' at the origin $0\in\R^m$ and
``target'' in~$G$, and can be given the structure of a (Lie) group. Indeed, let $S,T\in T^h_nG$
be any elements. We define a (smooth) multiplication in $T^h_mG$ by:
\begin{equation*}
T^h_m\mu\colon T^h_mG \times T^h_mG \to T^h_mG, \quad
(S=j^h_0a,T=j^h_0b)\mapsto S\cdot T:= j^{h}_{0}(a\cdot b),
\end{equation*}
where $(a\cdot b)(x):=a(x)\cdot b(x)\equiv\mu(a(x),b(x))$
is the group multiplication in~$G$.
The mapping $(S,T)\mapsto S\cdot T$ is associative;
moreover, the element $j^h_0e$, $e$ denoting both the unit element in~$G$
and the constant mapping from~$\R^n$ onto~$e$,
is the unit element of $T^h_nG$,
and $j^h_0a^{-1}$, where $a^{-1}(x):=\big(a(x)\big)^{-1}$
(the inversion being taken in the group $G$),
is the inverse of~$j^h_0a$.

\begin{dfn}
Consider a principal bundle $P(M,G)$.
Let $k$ and $h$ be two natural numbers such that $k\geq h$.
Then, by the \emph{$(k,h)$-principal prolongation} of~$P$ we shall mean the bundle
\begin{equation}
W^{k,h}\!P:=L^k\!M\ftimes{M} J^h\!P,
\label{eq:WP}
\end{equation}
$J^h\!P$ denoting the $h$-th order jet prolongation of~$P$.
A point of $W^{k,h}\!P$ is of the form $(j^k_0\epsilon,j^h_x\sigma)$, where
$\epsilon\colon\R^{m} \to M$ is locally invertible with $\epsilon(0)=x$, and
$\sigma\colon M\to P$ is a local section around the point $x\in M$.
\end{dfn}

\noindent
Unlike $J^h\!P$, $W^{k,h}\!P$ is a principal
bundle over~$M$, and its structure group is
\begin{equation*}
W^{k,h}_mG:= G^k_m \rtimes T^h_mG.
\end{equation*}
$W^{k,h}_mG$ is called the \emph{$(m;k,h)$-principal prolongation}
of~$G$. The group
multiplication on
$W^{k,h}_mG$ is defined by the following rule:
\begin{equation*}
(j^k_0\alpha,j^h_0a)\odot(j^k_0\beta,j^h_0b):=
\Big(j^k_0(\alpha\circ\beta),j^h_0\big((a\circ\beta)\cdot b\big)\Big),
\end{equation*}
`$\cdot$' denoting the group multiplication in~$G$.
The right action of $W^{k,h}_mG$ on $W^{k,h}\!P$
is then defined by:
\begin{equation*}
(j^k_0\epsilon,j^h_x\sigma)\odot(j^k_0\alpha,j^h_0a):=
\Big(j^k_0(\epsilon\circ\alpha),
j^h_x\big(\sigma\cdot(a\circ \alpha^{-1}\circ\epsilon^{-1})\big)\Big),
\end{equation*}
`$\cdot$' denoting now the canonical right action of~$G$ on~$P$.

\bigskip
In the case $h=0$, we have a direct product of Lie groups $W^{k,0}_mG:= G^k_m \times G$ and the
usual fibred product $W^{k,0}P\equiv L^k\!M\ftimes{M}P$ of principal bundles.

\begin{rem}
When $h=k$, the functor $W^{k,k}$ is often simply written $W^k$.
\end{rem}

\begin{dfn}
Let $\Phi\colon P\to P$ be an automorphism over a diffeomorphism
$\varphi\colon M\to M$ (\cf\ \S\ref{ssec:pbm}).
We define an \emph{automorphism} of $W^{k,h}\!P$ associated with~$\Phi$ by
\begin{equation}
W^{k,h}\Phi\colon W^{k,h}\!P\to W^{k,h}\!P, \quad
(j^k_0\epsilon,j^h_x\sigma)\mapsto
\big(j^k_0(\varphi\circ\epsilon),j^h_x(\Phi\circ\sigma\circ \varphi^{-1})\big).
\label{eq:WPhi}
\end{equation}
\end{dfn}

\begin{prop}
The bundle morphism $W^{k,h}\Phi$ preserves the right action\textup, thereby being a principal
automorphism\textup.
\end{prop}
By virtue of~\eqref{eq:WP} and~\eqref{eq:WPhi} $W^{k,h}$ turns out to be a functor from the category	of
principal \Gdash bundles over \mdash dimensional manifolds and local isomorphisms to the category of
principal \WGdash bundles \cite{kms93}. Now, let $P_\lambda:=W^{k,h}\!P\times_{\lambda} F$ be a
fibre bundle  associated with $P(M,G)$ via an action~$\lambda$ of $W^{k,h}_mG$ on a
manifold~$F$. There exists a canonical representation of the automorphisms of~$P$ induced
by~\eqref{eq:WPhi}. Indeed, if
$\Phi\colon P\to P$ is an automorphism over a diffeomorphism $\varphi\colon M\to M$, then we can define the
corresponding 
\emph{induced automorphism}~$\Phi_\lambda$ as
\begin{equation}
\Phi_\l\colon P_\l\to P_\l,\quad
\Phi_\l\colon[u,f]_\l\mapsto[W^{k,h}\Phi(u),f]_\l
\label{eq:indaut}
\end{equation}
which is well-defined since it turns out to be independent of the representative $(u,f)$, $u\in P$, $f\in
F$. This construction yields a functor $\cdot_\l$ from the category of principal \Gdash bundles to the
category of fibred manifolds and fibre-respecting mappings.
\begin{dfn}
A \emph{gauge-natural bundle of order~$(k,h)$} over~$M$ associated with $P(M,G)$ is any such functor\textup.
\end{dfn}
If we now restrict attention to the case $G=\{e\}$ and $h=0$, we can recover the classical notion of natural
bundles over~$M$. In particular, we have the following
\begin{dfn}
Let $\varphi\colon M\to M$ be a diffeomorphism\textup. We define an automorphism of $L^k\!M$
associated with~$\varphi$\textup, called its \emph{natural lift \textup(onto~$L^k\!M$\textup)}\textup, by
\begin{equation*}
L^k\varphi\colon L^k\!M\to L^k\!M,\quad
L^k\varphi\colon j^k_0\epsilon\mapsto j^k_0(\varphi\circ\epsilon).
\end{equation*}
\end{dfn}
Then, $L^k$ turns out to be a functor from the category of \mdash dimensional manifolds and local
diffeomorphisms to the category of principal \Gkdash bundles. Now, given any fibre bundle
associated with $L^k\!M$ and any diffeomorphism on~$M$, we can define a corresponding induced automorphism
in the usual fashion. This construction yields a functor from the category of \mdash dimensional
manifolds to the category of fibred manifolds.
\begin{dfn}
A \emph{natural bundle of order~$k$} over~$M$ is any such functor\textup.
\end{dfn} 

\begin{rem}\label{rem:LM}
In the sequel we shall always assume that $L^k\!M$ is equipped with the principal bundle
structure \emph{naturally} induced by the differentiable structure of the base manifold~$M$,
i.e.\ that $L^k\!M$ itself is a natural bundle over~$M$. This is, of course, possible because we
can always identify a principal bundle $P(M,G)$ with its associated bundle~$P_\l:=P\times_\l G$,
where $\l$ is the left action of~$G$ on itself.
\end{rem}

Now, if $\Xi$ is a \Gdash invariant vector field on~$P$ generating a
one-parameter group $\{\Phi_{t}\}$ of automorphisms of~$P$ and projecting on a vector
field~$\xi$ on~$M$ (\cf\ \S\ref{ssec:pbm}), we can define the \emph{induced vector
field}~$\Xi_\l$ on~$P_\lambda$ simply by setting
\begin{equation*}
\Xi_\l := \left.\frac\de{\de t}(\Phi_t)_\l\right\rvert_{t=0},
\end{equation*}
which obviously projects on the same vector field~$\xi$ on~$M$. Now, on relying on the general
theory of Lie derivatives \cite{trautman72,jk82,kms93,gm02}, we can give the following
\begin{dfn}\label{dfn:gnld}
Let $P_\l$ be a gauge-natural bundle associated with some principal bundle $P(M,G)$, $\Xi$ a
\Gdash invariant vector field on~$P$ projecting on a vector field~$\xi$ on~$M$, and
$\sigma\colon M\to P_\l$ a section of~$P_\l$. Then, by the
\emph{generalized} (\emph{gauge-natural}) \emph{Lie derivative} of~$\sigma$ with respect to~$\Xi$
we shall mean the map
\begin{equation}
£_\Xi\sigma\colon M\to V\!P_\l, \quad
£_\Xi\sigma :=T\sigma\circ\xi -\Xi_\l\circ\sigma,
\label{eq:gnld}
\end{equation}
$V\!P_\l$ denoting the vertical (tangent) bundle of~$P_\l$.
\end{dfn}
Definition~\ref{dfn:gnld} is the conceptually natural generalization of the classical notion of a
Lie derivative \cite{yano57}, to which it suitably reduces when applied to natural objects. The
``formal'' counterpart of~$£_\Xi\sigma$ is defined in the usual way.
\begin{dfn}\label{dfn:dl-f}
Let $P_\l$, $\Xi$ and $\sigma$ be as above. We call \emph{formal generalized Lie derivative} the
global bundle morphism $£_\Xi y\colon J^1\!P_\l\to V\!P_\l$ intrinsically defined as
\begin{equation*}
£_\Xi y\circ j^1\sigma =£_\Xi\sigma,
\end{equation*}
where $ £_\Xi\sigma$ is the Lie derivative of~$\sigma$ in the sense of Definition~\ref{dfn:gnld}.
\end{dfn}
Of course, if $P_\l$ is a gauge-natural \emph{vector} or \emph{affine} bundle (which is always
the case in classical field theory), then its vertical bundle splits as $V\!P_\l\cong P_\l\times_M
P_\l$ or $V\!P_\l\cong P_\l\times_M\vec P_\l$, respectively, $\vec P_\l$ being the vector bundle
on which the affine bundle~$P_\l$ is modelled. In this case, we can give the following
\begin{dfn}\label{dfn:rgnld}
Let $P_\l$ be a gauge-natural vector or affine bundle associated with some principal bundle
$P(M,G)$, and let~$\Xi$ and~$\sigma$ be as in Definition~\ref{dfn:gnld}. Then,
by the (\emph{restricted gauge-natural}) \emph{Lie derivative} of~$\sigma$ with
respect to~$\Xi$ we shall mean the second component of~$£_\Xi\sigma$. Analogously, we shall call
\emph{formal} (\emph{restricted}) \emph{Lie derivative} the second component of~$£_\Xi y$.
\end{dfn}
In the sequel, we shall not formally distinguish between generalized and restricted
Lie derivatives, as it should be clear from the context which is the operator under actual
consideration.

\section{Noether theorem and conserved quantities}\label{sec:nt}
\label{sec:pc}

Let $M$ be an \mdash dimensional (orientable) manifold. By a \emph{$k$-th order Lagrangian}
on a gauge-natural bundle~$P_\l$, associated with some 
principal bundle $P(M,G)$, we shall mean a base-preserving morphism
\begin{equation}
\L \colon J^k\!P_\l\to\A^mT^{*}\!M
\label{eq:lag}
\end{equation}
or, equivalently, a horizontal \mdash form $\L\in\Omega^m_0(J^k\!P_\l)$. Locally, $\L(j^k y)$
reads
\begin{equation*}
\L(x^\l,y^\ag\ida_{\bs\mu}) = L(x^\l,y^\ag\ida_{\bs\mu})\,\d s
\end{equation*}
for some scalar density~$L$ on $J^k\!P_\l$, $\d s\equiv \d x^{0}\wedge \d
x^{1}\wedge\dots\wedge\d x^{m-1}$ denoting the local volume element on~$M$.
Then we have the following result (\cf, e.g., \cite{kms93}).
\begin{prop}
There exist a \emph{global} morphism $\FF(\L)\colon J^{2k-1}\!P_\l\to
V^*\!J^{k-1}\!P_\l\otimes\A^{m-1}T^*\!M$ and a \emph{unique global} morphism $\E(\L)\colon
J^{2k}\!P_\l\to V^*\!P_\l\otimes\A^mT^*\!M$ such that
\begin{equation}
(\pi^{2k}_k)^\ast\dual(\d\L, J^k\Ups) = \dual(\E(\L),\Ups) +\dH\dual(\FF(\L),J^{k-1}\Ups)
\label{eq:fvf}
\end{equation}
for any vertical vector field~$\Ups$ on~$P_\l$\textup.\footnote{The use of the horizontal differential
in~\eqref{eq:fvf} is justified by the one-to-one correspondence between horizontal forms and base-preserving
morphisms mentioned in \S\ref{ssec:jets}.} Locally\textup,
\begin{equation*}
\E(\L)  = \(\frac{\de L}{\de y^\ag} 
           +\sum_{1\leq\lvert\bs\mu\rvert\leq k}(-1)^{\lvert\bs\mu\rvert}\, 
            \d_{\bs\mu}\frac{\de L}{\de y^\ag\ida_{\bs\mu}}\)\bar\d y^\ag\otimes\d s,
\end{equation*}
$\{\bar\d y^\ag\}$ denoting the fibre basis of $V^*\!P_\l:=(V\!P_\l)^*$ defined by requiring 
$\dual(\bar\d y^\ag,\de_\bg) =\delta^\ag\ida_\bg$\textup.
\end{prop}
Classically, $(\pi^{2k}_k)^\ast\dual(\d\L, J^k\Ups)$ is denoted by $\delta\L$, a notation we shall
use ourselves when there is no danger of confusion, and identity~\eqref{eq:fvf} is known as the
\emph{first variation formula}. $\E(\L)$ is called the \emph{Euler-Lagrange morphism}, while
$\FF(\L)$ is known as the \emph{\pc\ morphism}. It is not uniquely determined and depends in
general on a linear connection on~$M$. In the case $k=1,2$, though, this dependence disappears
and $\FF(\L)$ locally reads
\begin{equation*}
\FF(\L)  = \[\(\frac{\de L}{\de y^\ag\ida_\mu}
               -(k-1)\d_\nu\frac{\de L}{\de y^\ag\ida_{\mu\nu}}\)\bar\d y^\ag
             +(k-1)\frac{\de L}{\de y^\ag\ida_{\mu\nu}}\bar\d y^\ag\ida_\nu\]\otimes\d s_\mu,
\end{equation*}
where $\d s_\mu$ stands for $\de_\mu\inn\d s$ and $\bar\d y^\ag\ida_\nu$ is defined by requiring
$\dual(\bar\d y^\ag\ida_\nu,\de_\bg\ida^\rho) =\delta^\ag\ida_\bg\delta^\rho\ida_\nu$.

\begin{dfn}\label{dfn:Lsymm}
A configuration bundle automorphism $(\varphi,\Phi_\l)$, i.e.\ an automorphism~$\Phi_\l$
of~$P_\l$ induced by an automorphism~$\Phi$ of~$P$ over a diffeomorphism~$\varphi\colon M\to M$,
is called a (\emph{Lagrangian}) \emph{symmetry} for Lagrangian~\eqref{eq:lag} if it leaves~$\L$
unchanged, i.e.\ if $\A^mT^*\!\varphi\circ\L\circ j^1\Phi_\l=\L$.
\end{dfn}

Clearly, the infinitesimal version of Definition~\ref{dfn:Lsymm} is the following.

\begin{dfn}\label{dfn:iLsymm}
A vector field~$\Xi_\l$ generating a one-parameter group~$\{(\Phi_t)_\l\}$ of symmetries is 
called an \emph{infinitesimal} (\emph{Lagrangian}) \emph{symmetry}.
\end{dfn}

\begin{dfn}
Let $\Aut(P_\l)$ denote the group of all induced automorphisms of~$P_\l$. We shall say that a 
$k$-th order Lagrangian on~$P_\l$ is \emph{$\Aut(P_\l)$-invariant} if any induced automorphism
of~$P_\l$ is a symmetry (and any induced vector field on~$P_\l$ is an infinitesimal symmetry).
\end{dfn}

\begin{dfn}
A \emph{$k$-th order Lagrangian field theory on a gauge-natural bundle~$P_\l$} is a
field theory where the fields are represented by (local) sections of~$P_\l$ and the
equations of motion  can be formally written as
\begin{equation}
\E(\L)\circ j^{2k}\sigma = 0
\label{eq:el}
\end{equation}
for some suitable $\Aut(P_\l)$-invariant $k$-th order Lagrangian~$\L$ on~$P_\l$ and some section
$\sigma\in\Ga(P_\l)$. $P_\l$ is called the \emph{configuration bundle} of the theory, $P$ its
\emph{structure bundle} and $\sigma$ a \emph{critical section} of~$P_\l$, whereas (the local
expression of)  equation~\eqref{eq:el} is known as the \emph{Euler-Lagrange equations}. Whenever
an identity holds only modulo equation~\eqref{eq:el}, we shall say that it holds ``on shell''
and use the symbol `$\approx$' instead of the equals sign. In particular, we shall write
equation~\eqref{eq:el} itself simply as $\E(\L) \approx 0$.
\end{dfn}

All known classical Lagrangian field theories such as classical mechanics, elasticity,
gravitational field theories (including, in particular, Einstein's general relativity and the
Einstein-Cartan theory), electromagnetism, the Yang-Mills theory, bosonic and fermionic matter
field theories, topological field theories---as well as all their possible mutual
couplings---are Lagrangian field theories on some suitable gauge-natural (vector or affine)
bundle
\cite{eck81,kms93,fatibene99}.
\begin{prop}
Let $\Xi_\l$ be a vector field on~$P_\l$ induced by a \Gdash invariant vector field~$\Xi$
on~$P$ projecting on a vector field~$\xi$ on~$M$\textup, and $\L$ an $\Aut(P_\l)$-invariant
$k$-th order Lagrangian on~$P_\l$\textup. Then\textup,
\begin{equation}
\dual(\d\L, J^k\!£_\Xi y) =\dH(\xi\inn\L).
\label{eq:fi}
\end{equation}
\end{prop}
\begin{proof}
The result readily follows from Definition~\ref{dfn:iLsymm} and the properties of the formal Lie
derivative, taking into account the isomorphism $J^kV\!P_\l\cong VJ^k\!P_\l$.
\end{proof}
Identity~\eqref{eq:fi} is known as the \emph{fundamental identity}. Combining~\eqref{eq:fvf} and
\eqref{eq:fi} we get
\begin{equation}
\dH E(\L,\Xi) = W(\L,\Xi),
\label{eq:nt}
\end{equation}
where we set
\begin{align}
E(\L,\Xi) &:= \dual(\FF(\L),J^{k-1}\!£_\Xi y) -\xi\inn\L
\label{eq:E}  \\
\intertext{and}
W(\L,\Xi) &:= -\dual(\E(\L),£_\Xi y).
\notag
\end{align}
$E(\L,\Xi)$ is called the \emph{Noether current} and $W(\L,\Xi)$ the \emph{work form}. 
Formula~\eqref{eq:nt} is the generalization of \emph{Noether's theorem} \cite{noether18} to the
geometric framework of jet prolongations of gauge-natural bundles. Indeed, if we define
\begin{align*}
E(\L,\Xi,\sigma) &:= (j^{2k-1}\sigma)^\ast E(\L,\Xi),  \\
W(\L,\Xi,\sigma) &:= (j^{2k}\sigma)^\ast W(\L,\Xi),
\end{align*}
we have
\begin{equation*}
\d E(\L,\Xi,\sigma) = W(\L,\Xi,\sigma)
\end{equation*}
and, whenever~$\sigma$ is a critical section,
\begin{equation}
\d E(\L,\Xi,\sigma) = 0.
\label{eq:dE0}
\end{equation}
Thus, given an infinitesimal Lagrangian symmetry~$\Xi$, we have a whole class of currents
$E(\L,\Xi,\sigma)$ (one for each solution~$\sigma$), which are closed $(m-1)$-forms on~$M$. We
stress that the Noether current~$E(\L,\Xi)$ is defined at the bundle level and is
\emph{canonically} associated with the Lagrangian~$\L$. It is only at a \emph{later} stage that
it is evaluated on a section $\sigma\colon M\to P_\l$, thereby giving $E(\L,\Xi,\sigma)$.

Since $E(\L,\Xi,\sigma)$ is an $(m-1)$-form on~$M$, it can be integrated over an
$(m-1)$-dimensional region~$\Sigma$, namely a compact submanifold~$\Sigma\hookrightarrow M$ with
boundary~$\de\Sigma$.
\begin{dfn}
The real functional
\begin{equation}
Q_\Sigma(\L,\Xi,\sigma) := \int_\Sigma E(\L,\Xi,\sigma)
\label{eq:Q}
\end{equation}
is called the \emph{conserved quantity} (or \emph{charge}) along~$\sigma$ with respect to the
infinitesimal symmetry~$\Xi$ and the region~$\Sigma$.
\end{dfn}

\noindent
Indeed, if~$\sigma$ is a critical section, and two compact $(m-1)$-submanifolds $\Sigma,
\Sigma'\hookrightarrow M$ form the homological boundary~$\de D$ of a compact $m$-dimensional
domain $D\subseteq M$, from \eqref{eq:Q}, Stokes's theorem and~\eqref{eq:dE0} we readily obtain
\begin{equation*}
Q_{\Sigma'}(\L,\Xi,\sigma) = Q_\Sigma(\L,\Xi,\sigma).
\end{equation*}

Since $E(\L,\Xi,\sigma)$ is closed on shell, in field theories where $m>1$ we may ask ourselves
whether it is also exact, i.e.\ whether there exists an
$(m-2)$-form
$U(\L,\Xi,\sigma)$ on~$M$ such that
\begin{equation}
E(\L,\Xi,\sigma) = \d U(\L,\Xi,\sigma).
\label{eq:EdU}
\end{equation}
If this happens to be the case, then we can express $Q_\Sigma(\L,\Xi,\sigma)$ as an
$(m-2)$-dimensional integral over the boundary $\de\Sigma$ of~$\Sigma$. Indeed, on
using~\eqref{eq:EdU} and Stokes's theorem, we have
\begin{equation}
Q_\Sigma(\L,\Xi,\sigma) =\int_{\de\Sigma} U(\L,\Xi,\sigma).
   \label{eq:QdU}
\end{equation}
Actually, it is possible to prove the following fundamental 
\begin{thm}\label{thm:sp}
The Noether current is exact on shell for all gauge-natural field theories\textup, regardless
of the topology of~$M$
\textup{\cite{fatibene99}\footnotemark.}
\end{thm}
\footnotetext{The proof closely follows the analogous one for natural field theories due to
Robutti \cite{robutti84} (see also \cite{ffr86,ffr87}).}

We stress that this important result can only be achieved since Noether's theorem has been
formulated in terms of fibred morphisms rather than directly in terms of
forms on~$M$. Notably, we shall give the following

\begin{dfn}
If the Noether current $E(\L,\Xi)$ can be written as
\begin{equation}
E(\L,\Xi)=\tilde E(\L,\Xi) +\dH U(\L,\Xi),
\label{eq:EEdU}
\end{equation}
where $\tilde E(\L,\Xi,\sigma):=(j^{2k-1}\sigma)^\ast\tilde E(\L,\Xi)$ vanishes for any critical
section~$\sigma$, then we shall call $\tilde E(\L,\Xi)$ and $U(\L,\Xi)$ the \emph{reduced
current} and the \emph{superpotential} associated with~$\L$, respectively. Whenever the 
splitting~\eqref{eq:EEdU} holds, then it is immediate to see that
$U(\L,\Xi,\sigma):=(j^{2k-1}\sigma)^\ast U(\L,\Xi)$ satisfies equation~\eqref{eq:EdU} for any
critical section~$\sigma$.
\end{dfn}

Of course, neither Noether currents nor superpotentials are unique: the former are defined
modulo exact $(m-1)$-forms, the latter modulo closed $(m-2)$-forms. What is physically
meaningful, though, are the conserved quantities, which only depend on the cohomology class, not
on the chosen representative.

Finally, one might be interested in what happens to Noether currents and superpotentials (and,
hence, to the conserved quantities) when the Lagrangian of the theory is replaced by an
equivalent one. We recall that two Lagrangians~$\L$ and~$\L'$ are said to be \emph{equivalent}
if $\E(\L)=\E(\L')$. Owing to linearity, this is tantamount to saying that two Lagrangians are
equivalent if they differ from each other by a (\emph{variationally}) \emph{trivial} Lagrangian,
i.e.\ a Lagrangian whose Euler-Lagrange morphism is identically zero. The issue of finding all
trivial Lagrangians represents one of the most difficult problems of the geometric calculus of
variations and, in the $k$-th order case, was only recently solved by Krupka and Musilov\'a
\cite{km98}. In the present context, their result can be rephrased as follows.

\begin{thm}
Two $k$-th order Lagrangians~$\L$ and~$\L'$ on a gauge-natural bundle~$P_\l$ are equivalent
iff\textup, locally\textup, they differ from each other by the horizontal part of the exterior
differential of an $(m-1)$-form~$\chi$ on~$J^{k-1}\!P_\l$\textup.
\end{thm}

Therefore, if $\L':=\L+h(\d\chi)$ and we set $\beta:=h(\chi)$, we readily find
\begin{equation*}
E(\L',\Xi) = E(\L,\Xi) + £_{J^k\Xi}\beta -\xi\inn\dH\beta.
\end{equation*}
But $\beta\in\Omega^{m-1}_0(J^k\!P_\l)$; accordingly, $£_{J^k\Xi}\beta =£_\xi\beta \equiv
\xi\inn\dH\beta +\dH(\xi\inn\beta)$, whence
\begin{align}
\tilde E(\L',\Xi) &= \tilde E(\L,\Xi),  \notag \\
       U(\L',\Xi) &= U(\L,\Xi) +\xi\inn\beta.
\label{eq:UL'}
\end{align}

\section{Spin-tetrads, spin-connections and spinors}
\label{sec:st}

To the best of our knowledge, the concept of a ``free spin structure'' was originally introduced
(with a different purpose) in \cite{pw87} (see also \cite{swift88}). It was then rediscovered in
\cite{vdheuvel94} for the very reason mentioned in the Introduction and further analysed
in~\cite{fffg98,fafr98}. The notion of a ``spin-tetrad'' as a section of a suitable gauge-natural
bundle over~$M$ was first proposed in \cite{fffg98}.
\begin{dfn}\label{dfn:fss}
Let $M$ be a 4-dimensional manifold admitting Lorentzian metrics of signature $-2$, i.e.\
satisfying the topological requirements which ensure  the existence on it of  Lorentzian
structures [\SOdash reductions], and let $\Lambda$ be the epimorphism which exhibits
$\Spin(1,3)^e$  as the twofold  covering of $\SO(1,3)^e$. A \emph{free spin structure}
on~$M$ consists of a principal bundle $\pi\colon\Sigma\to M$ with structure group
$\Spin(1,3)^e$ and a map $\tilde\Lambda\colon\Sigma\to LM$ such that
\begin{gather*}
\tilde\Lambda\circ r^S =r^{\prime(\iota\circ\Lambda)(S)}\circ\tilde\Lambda 
                               \quad\forall S\in\Spin(1,3)^e,  \\
\pi'\circ\tilde\Lambda=\pi,
\end{gather*}
$r$ and $r'$ denoting the canonical right actions on~$\Sigma$ and~$LM$, respectively,
$\iota\colon\SO(1,3)^e\to\GL(4,\R)$ the canonical injection of Lie groups, and $\pi'\colon LM\to
M$ the canonical projection. We shall call the bundle map $\tilde\Lambda$ a \emph{spin-frame}
on~$\Sigma$.
\end{dfn}
\noindent
This definition of a spin structure induces metrics on $M$.
Indeed, given a spin-frame $\tilde\Lambda\colon\Sigma\to  LM$,
we can define a metric~$g$ via the reduced subbundle $\SO(M,g):=
\tilde\Lambda(\Sigma)$ of $ LM$. In other words, the (\emph{dynamic}) metric $g\equiv
g_{\tilde\Lambda}$ is defined to be the metric such that
frames in $\tilde\Lambda(\Sigma)\subset  LM$ are $g$-orthonormal frames. 
It is important to stress that in our picture the metric~$g$
is built up \emph{a posteriori}, after a spin-frame has been determined by the 
field equations in a way which is compatible with the (free) spin structure
one has used to define spinors.

Now, if we want to regard spin-frames as dynamical variables in a Lagrangian field theory, we
should be able to represent them as (global) sections of a suitable configuration bundle. This
motivates the following
\begin{dfn}\label{dfn:st}
Let $\Lambda$ be as in Definition~\ref{dfn:fss} and consider the following left action of the
group $W^{1,0}_4\Spin(1,3)^e$ on the manifold $\GL(4,\R)$
\begin{equation*}
\rho\colon\bigl((A,S),\theta\bigr)
\mapsto \theta' := \Lambda(S)\circ\theta\circ A^{-1}
\end{equation*}
together with the associated  bundle $\Sigma_\rho:=W^{1,0}\Sigma
\times_{\rho}\GL(4,\R)$.
$\Sigma_\rho$ is a fibre bundle associated with $W^{1,0}\Sigma$, 
i.e.\ a gauge-natural bundle of order~$(1,0)$. A section
of $\Sigma_\rho$ will be called a \emph{spin-tetrad}.
\end{dfn}
\noindent
If $(\theta^a\ida_\mu)$ denote the 
components of a spin-tetrad~$\theta$ in some local chart, then the components $(g_{\mu\nu})$ of
the induced metric~$g$ in the associated chart read
\begin{equation}
g_{\mu\nu} \equiv\theta^a\ida_\mu\theta^b\ida_\nu\eta_{ab},
\label{eq:gtt2}
\end{equation}
formally identical with equation~\eqref{eq:gtt} of the Introduction, but with
\emph{both~\eqref{eq:tet-tn} and~\eqref{eq:tet-tg}} built in.

Recall now that a (principal) connection on a principal 
bundle $P(M,G)$ may be regarded as a $G$-equivariant global 
section of the affine jet bundle $J^{1}\!P\to P$, where the $G$-action on $J^1\!P$
is induced by the first jet prolongation of the canonical
(right) action of $G$ on~$P$. Owing to $G$-equivariance there is a one-to-one correspondence 
between principal connections and global sections of the quotient bundle 
$J^{1}\!P/G\to M$ (\cf\ \cite{kms93,gms97}). Bearing this in mind, we can give
\begin{dfn}\label{dfn:sc}
Let $\so(1,3)\cong\spin(1,3)$ denote the Lie algebra of $\SO(1,3)^e$ and consider the
following left action of the group $W^{1,1}_4\Spin(1,3)^e
\cong\GL(4,\R)\times\Spin(1,3)^e\rtimes\bigl((\R^4)^*\otimes\so(1,3)\bigr)$ 
on the vector space $(\R^4)^*\otimes\so(1,3)$
\begin{equation*}
\ell\colon\bigl((A,S,\hat S),\omega\bigr)
         \mapsto \omega' := \Ad_{\Lambda(S)}\bigl(\omega -\hat\Lambda(\hat S)\bigr)\circ A^{-1},
\end{equation*}
where $\Ad$ denotes the adjoint representation of $\SO(1,3)^e$, and $\hat\Lambda
:=\id\otimes T_e\Lambda$ the isomorphism between $(\R^4)^*\otimes\spin(1,3)$ and
$(\R^4)^*\otimes\so(1,3)$ induced by~$\Lambda$. Clearly, the associated bundle
$\Sigma_\ell := W^{1,1}\Sigma\times_{\ell}\bigl((\R^4)^*\otimes\so(1,3)\bigr)$ is a
gauge-natural bundle of order~$(1,1)$ isomorphic to $J^1(\Sigma/\Z_2)/\SO(1,3)^e$. 
A section of~$\Sigma_\ell$ will be called a \emph{spin-connection}.
\end{dfn}

Note that also spinors can be regarded as sections of a suitable gauge-natural bundle over~$M$.
Indeed, if $\hga$ is the linear representation of $\Spin(1,3)^e$ on the vector space $\C^4$
induced by the given choice of $\gamma$ matrices, then the associated vector bundle $\Sigma_\hga
:=\Sigma\times_\hga\C^4$ is a gauge-natural bundle of order $(0,0)$ whose sections represent
\emph{spinors} (or, more precisely, \emph{spin-vector fields}). Therefore, in spite of what is
usually believed, a Lie derivative for spinors (in the sense of Definition~\ref{dfn:gnld}) always
exists, \emph{no matter what} the vector field~$\xi$ on~$M$ is (\cf\ \cite{fffg96,gm02}).

Finally, for the sake of completeness, we shall recall that the \emph{Dirac adjoint}~$\bar\psi$
of a spinor~$\psi$ is defined as the adjoint of~$\psi$ with respect to the standard
\Spindash invariant scalar product on~$\C^4$ (see, e.g., \cite{bt88}). With our
conventions, $\bar\psi$ locally reads
\begin{equation*}
\bar\psi(x) =\psi^\dag (x)\ga_0
\end{equation*}
for all $x\in M$. Note also that, in our picture, the \emph{spinor connection}~$\tilde\omega$
corresponding to a given spin-connection~$\omega$ may be defined in terms of~$\omega$ as
\begin{equation*}
\tilde\omega := \bigl(\id\otimes(\Lambda')^{-1}\bigr)(\omega),
\end{equation*}
$\Lambda':=T_e\Lambda$ denoting the Lie algebra isomorphism between $\spin(1,3)$ and
$\so(1,3)$. On differentiating~\eqref{eq:sg1} and taking~\eqref{eq:sg2} into
account, we find that $(\Lambda')^{-1}(\mathfrak{l})$ is given by\footnote{Here and in the
sequel, $\so(1,3)$ is understood to be represented on the Lie algebra of $4\times 4$ real
matrices by means of its ``fundamental representation'', i.e.\ its lowest dimensional faithful
(linear) representation.}
\begin{equation*}
(\Lambda')^{-1}(\mathfrak{l})\equiv -\frac14\mathfrak{l}^{ab}\ga_{ab}
\end{equation*}
for all $\mathfrak{l}\equiv(\mathfrak{l}^a\ida_c=:\mathfrak{l}^{ab}\eta_{bc})\in\so(1,3)$.
Therefore, the components~$(\tilde\omega_\mu)$ of~$\tilde\omega$ read
\begin{equation}
\tilde\omega_\mu\equiv -\frac14\omega^{ab}\ida_\mu\ga_{ab},
\label{eq:omegat}
\end{equation}
$(\omega^a\ida_{c\mu}=:\omega^{ab}\ida_\mu\eta_{bc})$ denoting the components of~$\omega$.
Identity~\eqref{eq:omegat} is what is used in practice for evaluating the covariant derivative
of a spinor and its Dirac adjoint,
\begin{align*}
\cd_\mu\psi &:= \de_\mu\psi +\tilde\omega_\mu\psi,  \\
\cd_\mu\bar\psi &\hphantom:= \overline{\cd_\mu\psi}\equiv\de_\mu\bar\psi
-\bar\psi\tilde\omega_\mu.
\end{align*}
On making use of~\eqref{eq:gnld} we can now readily evaluate the Lie derivatives, with
respect to a \Spindash invariant vector field~$\Xi$ on~$\Sigma$, of a spin-tetrad, a
spin-connection, a spinor and its Dirac adjoint, which will locally read
\begin{subequations}
\begin{align}
£_\Xi\theta^a\ida_\mu &=\xi^\nu\de_\nu\theta^a\ida_\mu +\de_\mu\xi^\nu\theta^a\ida_\nu
-\Xi^a\ida_b\theta^b\ida_\mu,
\label{eq:gnldst}  \\
£_\Xi\omega^{a}\ida_{b\mu}
      &=\xi^{\nu}\de_{\nu}\omega^{a}\ida_{b\mu}
        +\de_{\mu}\xi^{\nu}\omega^{a}\ida_{b\nu} 
        + \omega^{a}\ida_{c\mu}\Xi^{c}\ida_{b} 
        - \omega^{c}\ida_{b\mu}\Xi^{a}\ida_{c}
        +\de_{\mu}\Xi^{a}\ida_{b},
\label{eq:gnldsc}  \\
£_\Xi\psi &=\xi^\nu\de_\nu\psi +\tfrac14\Xi^{ab}\ga_{ab}\psi,  \\  
£_\Xi\bar\psi &=\overline{£_\Xi\psi} \equiv\xi^\nu\de_\nu\bar\psi
-\tfrac14\Xi^{ab}\bar\psi\ga_{ab},
\end{align}
\end{subequations}
respectively, $(\xi^\mu, \Xi^a\ida_c=:\Xi^{ab}\eta_{bc})$ denoting the components of the
\SOdash invariant vector field $\Xi_{\tilde\Lambda}$ induced by~$\Xi$.

\section{Einstein (-Cartan) -Dirac theory}
\label{sec:ecd}

Throughout this section we shall use Cartan's language of vector (bundle)-valued differential
forms (on~$M$), which will prove to be an elegant and compact way to express our findings.
To this end, let $\Sigma_\hrho:=\Sigma\times_\hrho\R^4$ denote the vector bundle associated
with~$\Sigma$ via the action
\begin{equation*}
\hrho\colon\Spin(1,3)^e\times\R^4\to\R^4, \quad
(S,u)\mapsto\Lambda(S)\circ u.
\end{equation*}
Then, a spin-tetrad can be equivalently
regarded as a $\Sigma_\hrho$-valued \onedash form on~$M$ locally reading
\begin{equation}
\theta :=\theta^a\otimes f_a, \quad 
\theta^a:=\theta^a\ida_\mu\,\d x^\mu,
\label{eq:st-def2}
\end{equation}
$(f_a)$ denoting a local fibre basis of~$\Sigma_\hrho$. Furthermore, let $\gl(\Sigma_\hrho)$
denote the vector bundle over~$M$ given by the value at~$\Sigma_\hrho$ of the canonical
extension of the functor~$\gl$ to the category of vector bundles and their homomorphisms (see
\cite[\S6.7]{kms93}). Finally, if $(\omega^a\ida_{b\mu})$ are the components of a spin-connection
in some local chart, it is convenient to introduce the notation
\begin{equation*}
\omega^a\ida_b :=\omega^a\ida_{b\mu}\,\d x^\mu.
\label{eq:omega1f}
\end{equation*}

\subsection{Riemann-Cartan geometry on spin manifolds}

Now, let `$\nabla$' be the covariant derivative operator with respect to the connection
on~$T^p_qM$ \emph{naturally} induced by a linear connection~$\Gamma$ on~$LM$, $T^p_qM$ denoting
the $(p,q)$-tensor bundle over~$M$. Classically, Riemann-Cartan geometry is characterized by two
conditions: the covariant constancy of the metric,
\begin{equation}
\nabla g = 0,
\label{eq:metricity}
\end{equation}
just as in ordinary Riemannian geometry, and the presence of a (not necessarily zero) torsion
tensor~$\tau$ such that
\begin{equation*}
\tau(\xi,\xi') =\cd_\xi\xi' -\cd_{\xi'}\xi -[\xi,\xi']
\end{equation*}
for any two vector fields~$\xi$ and~$\xi'$ on~$M$.

In the present gauge-natural setting we can introduce analogous concepts serving a similar
purpose. In particular, if $\theta$ is a spin-tetrad in the sense of Definition~\ref{dfn:st}
and~$g$ is the metric induced by~$\theta$ via~\eqref{eq:gtt2}, equation~\eqref{eq:metricity}
can be derived by the condition 
\begin{equation}
\nabla\theta = 0,
\label{eq:cdtheta}
\end{equation}
where, here, `$\nabla$' denotes the covariant derivative operator with respect to the connection
on~$\Sigma_\rho$ \emph{canonically} induced by the connections~$\Gamma$ and~$\tilde\omega$
on~$LM$ and $\Sigma$, respectively (see also \S\ref{ssec:na}). Accordingly, we can define a
\emph{torsion \twodash form} as the $\Sigma_\hrho$-valued 2-form, which we shall denote again
by~$\tau$, given by the expression
\begin{equation*}
\tau :=\D\theta
\end{equation*}
or, equivalently,
\begin{equation}
\tau:=\tau^a\otimes f_a, \quad
\tau^a :=\D\theta^a \equiv\d\theta^a +\omega^a\ida_b\wedge\theta^b,
\label{eq:tau}
\end{equation}
`$\D$' denoting the ``covariant exterior derivative'' operator \cite[\S11.13 \emph{et
seq.}]{kms93} and~$\theta$ being as in~\eqref{eq:st-def2}.  Moreover, we can define a
\emph{contortion \onedash form} as the
$\gl(\Sigma_\hrho)$-valued 1-form measuring the deviation of the
spin-connection~$\omega$ from the Riemannian (or ``Levi-Civita'')
spin-connection~$\thomega$ [see \eqref{eq:thomega} below]:
\begin{equation*}
K := (\omega^a\ida_b -\thomega^a\ida_b)\otimes F_a\ida^b,
\end{equation*}
$(F_a\ida^b)$ denoting a local fibre basis of~$\gl(\Sigma_\hrho)$. The components of the
associated tensor field then read
\begin{equation}
K^{abc} = -\frac12(\tau^{abc}+\tau^{bca}-\tau^{cab}),
\label{eq:K}
\end{equation}
$(\tau^a\ida_{de}=:\tau^{abc}\eta_{bd}\eta_{ce})$ denoting
the components of the tensor associated to the torsion 2-form. Finally, note that a
\emph{curvature \twodash form} associated with~$\omega$ may be defined as the
$\gl(\Sigma_\hrho)$-valued 2-form
\begin{equation*}
\Omega :=\Omega^a\ida_b\otimes F_a\ida^b, \quad
\Omega^{a}\ida_{b} :=\d\omega^{a}\ida_{b} +\omega^{a}\ida_{c}\wedge\omega^{c}\ida_{b},
\end{equation*}
and that the components of the Riemannian spin-connection~$\thomega$ read (\cf~\cite{c-b87})
\begin{equation}
\thomega^{ab}\ida_\mu =\theta^{b\nu}\de_{[\nu}\theta^a\ida_{\mu]}
+\theta^{a\rho}\theta_{c\mu}\theta^{b\nu}\de_{[\nu}\theta^c\ida_{\rho]}
+\theta^{a\nu}\de_{[\mu}\theta^b\ida_{\nu]} \equiv\thomega^{[ab]}\ida_\mu ,
\label{eq:thomega}
\end{equation}
Latin and Greek indices being lowered or raised by~$\eta$ and~$g$, respectively, or their
inverses.

\bigskip
We are now in a position to apply the theory of conserved quantities developed in~\S\ref{sec:pc}
to the Einstein (-Cartan) -Dirac theory. We shall do so separately for the Einstein-Cartan-Dirac
case and the Einstein-Dirac one. Calculations will be ``formal'', unless otherwise stated, i.e.\
they will involve local coordinates, rather than sections, of the bundles under consideration.
For the sake of simplicity, we shall nevertheless use the names of the corresponding sections.
With a slight abuse of notation, we shall also use the symbols~`$\nabla$' and~`$\D$' for their
formal counterparts, defined in the usual manner (\cf, e.g., Definition~\ref{dfn:dl-f}).

\subsection{Einstein-Cartan-Dirac theory}
\label{ssec:ECD}

Our main reference for the Einstein-Cartan-Dirac theory is~\cite{c-b87}.

In the light of the new geometric framework developed in \S\ref{sec:st}, the
\emph{Einstein-Cartan Lagrangian} can be defined as the base-preserving morphism
\begin{equation*}
	\LH \colon \Sigma_{\rho}\ftimes{M}
	       J^{1}\Sigma_\ell\to\A^{4} T^{*}\!M ,\quad
       \LH (\theta,j^1\omega)
       := -\frac{1}{2\kappa}\Omega_{ab}\wedge\Sigma^{ab},
\end{equation*} 
where $\kappa:=8\pi G/c^4$, $\Sigma_{ab}:=e_b\inn(e_a\inn\Sigma)$ and $\Sigma$ is the
standard volume form on~$M$ locally given by $\det\lVert\theta\rVert\,\d x^0\wedge\dots
\wedge\d x^3$. Here $\lVert\theta\rVert$ stands for the matrix of the components of~$\theta$ and
we have set $e_a:=e_a\ida^\mu\de_\mu$, $\lVert e_a\ida^\mu\rVert$ denoting the inverse of
$\lVert\theta\rVert$. The \emph{Dirac Lagrangian} reads instead
\begin{equation*}
\LD \colon \Sigma_{\rho}\ftimes{M}\Sigma_{\ell}\ftimes{M}J^{1}\Sigma_{\hga}
 \to\A^{4}T^{*}\!M,\quad
\LD(\theta,\omega,j^1\psi) :=
\left[\frac{\mathrm{i}\alpha}2(\bar\psi\gamma^{a}\cd_{a}\psi
 -\cd_{a}\bar\psi\gamma^{a}\psi) -m\bar\psi\psi\right]\Sigma,
\end{equation*}
where $\alpha:=\hslash c$. According to the principle of minimal coupling, the total
Lagrangian of the theory will be simply assumed to be $\L:=\LH+\LD$. A vertical vector field on 
the configuration bundle will then read
\begin{equation*}
\Ups= \del\theta^a\ida_\mu\frac\de{\de\theta^a\ida_\mu} +\del\omega^a\ida_{b\mu}\frac\de{\de
\omega^a\ida_{b\mu}} +\del\omega^a\ida_{b\mu,\nu}\frac\de{\de\omega^a\ida_{b\mu,\nu}}
+\del\psi^A\frac\de{\de\psi^A} 
+\del\psi^A\ida_\mu\frac\de{\de\psi^A\ida_\mu},
\end{equation*}
where $(\theta^a\ida_\mu)$, $(\omega^a\ida_{b\mu},\omega^a\ida_{b\mu,\nu})$ and
$(\psi^A,\psi^A\ida_\mu)$ denote fibre coordinates on $\Sigma_\rho$, $J^1\Sigma_\ell$ and
$J^1\Sigma_\hga$, respectively. If we set locally
\begin{alignat*}{2}
\del\theta^a &:=\del\theta^a\ida_\mu\,\d x^\mu,&\qquad
\del\omega^a\ida_b &:=\del\omega^a\ida_{b\mu}\,\d x^\mu,  \\
\del\psi &:=\del\psi^A\,f_A,
&\del\bar\psi &:=\overline{\del\psi}, 
\end{alignat*}
$(f_A)$ denoting a local fibre basis of~$\Sigma_\hga$,  
then the first variation formula for~$\L$ is
\begin{multline}
\del\L =\(-\tfrac1\kappa G^a\ida_b +T^a\ida_b\)\Sigma_a\wedge\del\theta^b
+\(\tfrac1{2\kappa}\D\Sigma_{ab} -S_{ab}\ida^c\Sigma_c\)\wedge\del\omega^{ab}  \\
+\dH\[-\tfrac1{2\kappa}\Sigma_{ab}\wedge\del\omega^{ab}
-\tfrac{\mathrm{i}\alpha}{2}(\del\bar\psi\ga^a\psi -\bar\psi\ga^a\del\psi)\Sigma_a\]  \\
+\alpha[\del\bar\psi\,\E(\LD) +\bar\E(\LD)\,\del\psi]\Sigma,
\label{eq:fvfECD}
\end{multline}
where $G$ denotes the Einstein tensor associated with~$\Omega$, $\Sigma_a:=e_a\inn\Sigma$ and we
set
\begin{gather*}
\begin{alignat*}{2}
T^a\ida_b       &:= \Theta^a\ida_b -\tfrac\alpha2[\bar\psi\,\thE(\LD)
                     +\thbE(\LD)\,\psi]\del^a\ida_b, &\qquad
\Theta^a\ida_b  &:= \tfrac{\mathrm{i}\alpha}2(\bar\psi\ga^a\cd_b\psi -\cd_b\bar\psi\ga^a\psi), \\
\thE(\LD)      &:= \mathrm{i}\ga^a\cd_a\psi -m\psi, &\qquad 
\thbE(\LD)     &:= \overline{\thE(\LD)} \equiv -(\mathrm{i}\cd_a\bar\psi\ga^a +m\bar\psi),  \\
\E(\LD)        &:= \thE(\LD)-\tfrac{\mathrm{i}}2K^a\ida_{ba}\ga^b\psi, &\qquad 
\bar\E(\LD)    &:= \overline{\E(\LD)} \equiv
\thbE(\LD)+\tfrac{\mathrm{i}}2K^a\ida_{ba}\bar\psi\ga^b,
\end{alignat*} \\
S^{abc}         := -\tfrac{\mathrm{i}\alpha}{8}\bar\psi(\ga^{ab}\ga^c +\ga^c\ga^{ab})\psi
                  \equiv -\tfrac{\mathrm{i}\alpha}{4}\bar\psi\ga^{abc}\psi \equiv S^{[abc]},
\end{gather*}
identity~\eqref{eq:gabc} having been used in the last but one equality.
Thus, the Einstein-Cartan-Dirac equations are
\begin{gather*}
G_{ab} \approx \kappa T_{ab},  \\
\D\Sigma_{ab} \approx 2\kappa S_{ab}\ida^c\Sigma_c, \\
\mathrm{i}\ga^a\cd_a\psi -m\psi -\tfrac{\mathrm{i}}2K^a\ida_{ba}\ga^b\psi \approx 0.
\end{gather*}
The first two equations are called the \emph{first} and the \emph{second Einstein-Cartan
\emph(\nobreakdash-\hspace{0pt}Dirac\emph) equation}, respectively, whereas the last one is
known as the \emph{\emph(Cartan\nobreakdash-\emph) Dirac} equation. $T$ is the
\emph{energy-momentum tensor} of the theory, and~$S$ the \emph{spin momentum tensor}. Now, making
use of~\eqref{eq:tau}, the second Einstein-Cartan equation can be put into the form
\begin{equation*}
\tau^c\wedge\Sigma_{abc} \approx 2\kappa S_{ab}\ida^c\Sigma_c
\end{equation*}
or equivalently
\begin{equation}
\tau^{abc} \approx 2\kappa S^{abc},
\label{eq:EC2}
\end{equation}
which in turn implies that the torsion tensor is completely antisymmetric on shell.
Therefore, so is the contortion tensor. Indeed, from~\eqref{eq:K} and~\eqref{eq:EC2}
\begin{equation*}
K^{abc} \approx -\frac12\tau^{abc} \approx -\kappa S^{abc}.
\end{equation*}
Hence, the Dirac equation reduces to $\thE(\LD)\approx0$, which implies
$T_{ab}\approx\Theta_{ab}$. To sum up, the above system of equations is completely equivalent to
the following
\begin{gather*}
G_{ab} \approx \kappa\Theta_{ab},  \\
\tau^{abc} \approx 2\kappa S^{abc}, \\
\mathrm{i}\ga^a\cd_a\psi -m\psi \approx 0.
\end{gather*}

Comparison between~\eqref{eq:fvf} and~\eqref{eq:E} tells us that we can read off the Noether
current associated with~$\L$ from its first variation~\eqref{eq:fvfECD}:
\begin{equation*}
E(\L,\Xi) = -\xi\inn\L -\frac1{2\kappa}\Sigma_{ab}\wedge£_\Xi\omega^{ab} 
                  -\frac{\mathrm{i}\alpha}{2}(£_\Xi\bar\psi\ga^a\psi 
                  -\bar\psi\ga^a£_\Xi\psi)\Sigma_a.
\end{equation*}
After some manipulation, which makes use (\emph{inter alia}) of the fact that the (formal) Lie 
derivative~\eqref{eq:gnldsc} of~$\omega$ can be put into the form
\begin{equation}
    £_{\Xi}\omega^{a}\ida_{b}=\xi\inn\Omega^{a}\ida_{b}+\D\VXi^{a}\ida_{b},
\label{eq:gnldsc2}
\end{equation}
$(\VXi^{a}\ida_{b}:=\Xi ^{a}\ida_{b}+\omega^{a}\ida_{b\mu}\xi^{\mu})$ 
being the components of the vertical part of~$\Xi_{\tilde\Lambda}$ with respect to~$\omega$,
and that $G^a\ida_b\Sigma_a \equiv -1/2\,\Omega^{ac}\wedge(e_b\inn\Sigma_{ac})$, $E(\L,\Xi)$ can
be recast as
\begin{equation*}
E(\L,\Xi) = \xi^b\(-\frac1\kappa G^a\ida_b +T^a\ida_b\)\Sigma_a 
             +\VXi^{ab}\(\frac1{2\kappa}\D\Sigma_{ab} -S_{ab}\ida^c\Sigma_c\)
             +\dH\(-\frac1{2\kappa}\VXi^{ab}\Sigma_{ab}\),
\end{equation*}
so that the superpotential associated with~$\L$ turns out to be
\begin{equation}
    U(\L,\Xi) := -\frac1{2\kappa}\VXi^{ab}\Sigma_{ab},
    \label{eq:ECDsp}
\end{equation}
a result which appeared in~\cite{gmv01} for the first time. Therefore, the Dirac Lagrangian
does not seem to contribute to the total superpotential. From this fact one might mistakenly
conclude that the Dirac fields do not contribute to the total conserved quantities. This
conclusion would be wrong because, although the Dirac Lagrangian does not contribute directly
to the superpotential, in order to obtain the corresponding conserved quantities, one needs
integrate the superpotential on a solution, which in turn depends on the Dirac Lagrangian via
its energy-momentum tensor and the second Einstein-Cartan equation.

Note that in the case of the ``Kosmann lift'' \cite{fffg96} (see also~\cite{gm02}) we have
\begin{equation}
    \VXi^{ab} = (\VxiK)^{ab} \equiv - \tilde\nabla^{[a}\xi^{b]},
    \label{eq:Klift}
\end{equation}
`$\tilde\nabla$' denoting the covariant derivative with respect to the transposed
connection. Substituting \eqref{eq:Klift} into~\eqref{eq:ECDsp} gives a half of the well-known
``Komar potential'' \cite{komar59}, in accordance with the result originally found by
Kijowski~\cite{kijowski78} (see also \cite{ffm94}) in a purely  natural context. This is
also the lift implicitly used in~\cite{fffg98,gmff00}.

\subsubsection{Natural approach}\label{ssec:na}
Suppose for a moment that we deliberately neglected the gauge-natural nature of the
Einstein-Cartan-Dirac theory. This means that we shall temporarily regard the Einstein-Cartan
Lagrangian as a purely natural Lagrangian, i.e.\ a first order Lagrangian on a (purely) natural
bundle. In particular, the spin-connection~$\omega$ will be replaced by a linear
connection~$\Gamma$, i.e.\ a principal connection on~$LM$, the latter regarded as a natural
bundle over~$M$ (\cf\ Remark~\ref{rem:LM}). As such, $\Gamma$ \emph{is} a natural object, whose
components~$(\Gamma^\rho\ida_{\nu\mu})$ are related, because of~\eqref{eq:cdtheta}, to the
components~$(\omega^a\ida_{b\mu})$ of~$\omega$ via the familiar formula\footnote{Note that
formula~\eqref{eq:omegaGamma} is \emph{formally} identical---not surprisingly---with the usual
expression for the components of the pull-back of~$\Gamma$ on~$M$ in an (anholonomic)
orthonormal basis. Note also that, for~$\Gamma^\rho\ida_{\nu\mu}$, we use a different subscript
ordering from~\cite{kn63}.}
\begin{equation}
\omega^a\ida_{b\mu} =\theta^a\ida_\rho(\de_\mu e_b\ida^\rho
+\Gamma^\rho\ida_{\nu\mu}e_b\ida^\nu),
\label{eq:omegaGamma}
\end{equation}
where antisymmetrization in~$\{a,b\}$ is understood on the right-hand side 
of~\eqref{eq:omegaGamma}.  Note that we can\emph{not} regard the Dirac Lagrangian itself as a
natural Lagrangian because spinors cannot be suitably replaced by any (physically equivalent)
natural objects: this is precisely why we went for a gauge-natural formulation in the first
place, and why we expect to encounter some sort of restrictions now.

The local expression for the Lie derivative of~$\Gamma$ reads
\begin{equation}
£_\xi\Gamma^\rho\ida_{\nu\mu} := 
£_\Xi\Gamma^\rho\ida_{\nu\mu}  =R^\rho\ida_{\nu\sigma\mu}\xi^\sigma 
                                +\cd_\mu\tilde\nabla\ida_\nu\xi^\rho,
\label{eq:ldGamma}
\end{equation}
a formula that has been known for a long time (\cf, e.g., \cite{schouten54,yano57}) and can be
evaluated directly from~\eqref{eq:gnld} or, equivalently, starting from~\eqref{eq:gnldsc2} and
then  using~\eqref{eq:omegaGamma} and~\eqref{eq:gnldst}. Thus, the Noether current is now of the
form
\begin{equation}
E(\L,\Xi) = -\xi\inn\L -\frac1{2\kappa}\Sigma_{ab}\wedge£_\xi\Gamma^{ab} 
                  -\frac{\mathrm{i}\alpha}{2}(£_\Xi\bar\psi\ga^a\psi 
                  -\bar\psi\ga^a£_\Xi\psi)\Sigma_a,
\label{eq:Ena}
\end{equation}
where we set 
\begin{equation*}
£_\xi\Gamma^{ab}:= (£_\xi\Gamma^\rho\ida_{\nu\mu})\theta^a\ida_\rho\theta^{b\nu}\,\d x^\mu.
\end{equation*}
The important point to note here is that, although $(\Gamma^\rho\ida_{\nu\mu})$ may be regarded
as the components of~$\omega$ in a holonomic basis, $(£_\xi\Gamma^\rho\ida_{\nu\mu})$ are
\emph{not}, in general, the components of~$£_\Xi\omega$ in the corresponding basis. Accordingly,
the second term on the right-hand side of identity~\eqref{eq:Ena} cannot be claimed to be  the
most general expression for $\dual(f(\LH),£_\Xi\omega)$, but naturality must indeed be assumed.
In fact, if we now proceeded in the same way as before, we would then find that consistency with
the second Einstein-Cartan equation requires $\VXi^{ab} = - \tilde\nabla^{[a}\xi^{b]}$, i.e.\
precisely the Kosmann lift, and thus we would recover the purely natural result, as expected.

\subsection{Einstein-Dirac theory}

Our main reference for the Einstein-Dirac theory is~\cite{lichnerowicz64}. The procedure for
obtaining the conserved quantities is completely analogous to the Einstein-Cartan-Dirac case;
therefore, we shall limit ourselves to present the results and briefly comment on them, pointing
out the possible differences. In the sequel, the symbol `$\vert_{K=0}$' affixed to a quantity
shall mean that the latter is formally identical with the quantity denoted by the same letter
in~\S\ref{ssec:ECD}, but with all (explicit or implicit) occurrences of~$\omega$ replaced
by~$\thomega$.

Then, the \emph{Einstein-Hilbert Lagrangian} is nothing but
\begin{equation*}
\thLH \colon J^{2}\Sigma_{\rho}\to\A^{4} T^{*}\!M,\quad
\thLH (j^2\theta):= \LH\rvert_{K=0},
\end{equation*} 
whereas the \emph{Dirac Lagrangian} is regarded here as the base-preserving morphism
\begin{equation*}
\thLD \colon J^1\Sigma_{\rho}\ftimes{M}J^1\Sigma_{\hga}
 \to\A^{4}T^{*}\!M,\quad \thLD(j^1\theta,j^1\psi):= \LD\rvert_{K=0}.
\end{equation*}
Again, the total Lagrangian of the theory will simply be $\thL:=\thLH+\thLD$, its
variation reading
\begin{multline*}
\del\,\thL =\(-\tfrac1\kappa\thG^a\ida_b +\thT^a\ida_b\)\Sigma_a\wedge\del\theta^b  \\
+\dH\[-\tfrac1{2\kappa}\Sigma_{ab}\wedge\del\,\thomega^{ab} +\tfrac12
S^{ab}\ida_c\Sigma_{ab}\wedge\del\theta^c -\tfrac{\mathrm{i}\alpha}{2}(\del\bar\psi\ga^a\psi
-\bar\psi\ga^a\del\psi)\Sigma_a\]  \\ +\alpha[\del\bar\psi\,\E(\thLD)
+\bar\E(\thLD)\,\del\psi]\Sigma,
\end{multline*}
where
\begin{gather}
\begin{align}
\thT^a\ida_b     &:=T^a\ida_b\rvert_{K=0} +\thcd_c S^a\ida_b\ida^c
                    \label{eq:thT1}  \\
                 &\hphantom{:}\equiv \thTheta^a\ida_b +\bar b^a\ida_b\,\E(\thLD)
                                  +\bar\E(\thLD)\,b^a\ida_b,
                     \label{eq:thT2}  \\
\thTheta^a\ida_b &:=\tfrac12(\Theta^a\ida_b +\Theta_b\ida^a)_{K=0}
                  \equiv \thTheta_b\ida^a,  \notag
\end{align}  \notag  \\
\begin{alignat*}{2}
b^a\ida_b        &:=\tfrac\alpha4(\ga^a\ida_b -2\del^a\ida_b)\psi,  &\qquad
\bar b^a\ida_b   &:=\overline{\hbox{$b^a\ida_b$}} 
                  \equiv -\tfrac\alpha4\bar\psi(\ga^a\ida_b +2\del^a\ida_b),  \\
\E(\thLD)        &:=\thE(\LD)\rvert_{K=0},  \notag  &\qquad
\bar\E(\thLD)    &:=\overline{\E(\thLD)} \equiv \thbE(\LD)\rvert_{K=0}.
\end{alignat*}  \notag
\end{gather}
Thus, the Einstein-Dirac equations are
\begin{gather*}
\thG_{ab} \approx \kappa \thTheta_{ab},  \\
\mathrm{i}\ga^a\,\thcd_a\psi -m\psi \approx 0.
\end{gather*}
Note that, although the invariance of the Dirac Lagrangian with respect to Lorentz
transformations requires $\thT_{ab}$ to be symmetric \emph{on} shell \cite{weinberg72,c-b87}, the
manipulation required for going from~\eqref{eq:thT1} to~\eqref{eq:thT2} is highly non-trivial:
the interested reader is referred to~\cite{lichnerowicz64} for an elegant proof.

Following the same procedure as before, we find that the Noether current associated with~$\thL$
is
\begin{align*}
E(\thL,\Xi) &= -\xi\inn\thL -\tfrac1{2\kappa}\Sigma_{ab}\wedge£_\Xi\thomega^{ab} 
+\tfrac12S^{ab}\ida_c\Sigma_{ab}\wedge£_\Xi\theta^c
-\tfrac{\mathrm{i}\alpha}{2}(£_\Xi\bar\psi\ga^a\psi
-\bar\psi\ga^a£_\Xi\psi)\Sigma_a  \\
          &= \xi^b\bigl(-\tfrac1\kappa\thG^a\ida_b +\thT^a\ida_b\bigr)\Sigma_a
             +\dH\bigl(-\tfrac1{2\kappa}\VXi^{ab}\Sigma_{ab}
                   +\tfrac12\xi^cS^{ab}\ida_c\Sigma_{ab}\bigr),
\end{align*}
so that the superpotential associated with~$\thL$ is recognized to be
\begin{equation}
    U(\thL,\Xi) := -\frac1{2\kappa}\VXi^{ab}\Sigma_{ab} +\frac12\xi^cS^{ab}\ida_c\Sigma_{ab},
    \label{eq:EDsp}
\end{equation}
and we note that, unlike the Einstein-Cartan-Dirac case, the Dirac Lagrangian enters the
superpotential directly, but recall that we have no second Einstein-Cartan equation here. Note
also that the ``vertical contribution'' (i.e.\ all terms in~$\VXi^{ab}$) coming from the Dirac
Lagrangian consistently vanishes \emph{off} shell. For the same reason, no inconsistency
of the type of~\S\ref{ssec:na} can arise here. This fact, though, by no means disproves the
gauge-naturality of the theory, which is well-motivated on both physical and mathematical
grounds.

\subsection{The indeterminacy}

Both~\eqref{eq:ECDsp} and~\eqref{eq:EDsp} reveal that, in this gauge-natural
formulation of gravity coupled with Dirac fields, the superpotential is essentially indeterminate
because no condition can be imposed \emph{a priori} on the vertical part of~$\Xi$. Therefore,
we can state our main result as follows.

\begin{thm}
Any conserved charge associated with the gravitational field is intrinsically
indeterminate\textup.
\end{thm}

Note that, because of~\eqref{eq:UL'}, this indeterminacy does not depend on the particular
Lagrangians chosen: for this reason and the functorial nature of this indeterminacy we have
called it ``intrinsic''. This important result can be regarded either as a limit for the theory
or as an additional flexibility (\cf~\cite{gmv01}). In any case, it cannot be overlooked.

From a physical point of view, it might be disturbing to think, that, when the spinorial
contribution is removed, the (gravitational part of the) theory should automatically revert to
its purely natural counterpart, thereby reproducing the well-known (non-indeterminate%
\footnote{But not always satisfactory: see \cite{gmff00}, \S4.}) results. This could mean either
that some justification has to be found to impose the Kosmann lift by hand or, conversely, as we
believe, that a gauge-natural formulation is the appropriate one for gravity for the very reason
that it is the most general one\footnote{Note that, even if we were to couple Einstein (-Cartan)
gravity with ``$U_4$-spinors'' \cite{buchdahl89,buchdahl92,godlewski02}, $\so(1,3)$ is in some
sense ``maximal'' since the superpotential of the theory must be a 2-form.}, irrespectively of
the nature of the theory it is possibly coupled with.

\section*{Acknowledgements}

The author would like to express his deep gratitude to Dr.~Marco Godina and Prof.~James~A.
Vickers for their useful remarks and encouragement. Also, this paper is, in many
respects, the climax of a long-term research project carried out over several years by the
mathematical physics group of the Department of Mathematics of the University of Turin, Italy,
whose indirect contribution to this work is hereby gratefully acknowledged.

\end{document}